\documentstyle[osa]{revtex}

\begin{document}

\title{
\begin{flushright}
{\small USACH-FM-00/07}\\[0.5cm]
\end{flushright}
{\bf Deformed Heisenberg Algebra with Reflection,
Anyons and Supersymmetry of 
Parabosons}\footnote{Talk 
given at the International Conference on 
``Spin-Statistics Connection and Commutation Relations", 
Centro Internazionale per la Cultura Scientifica dell'Universit\`a di 
Napoli, Anacapri, Capri Island, Italy -- May 31-June 3, 2000}}
 

\author{{\sf Mikhail S. Plyushchay}$^{\star,\dagger,\sharp}$\\
{\small {\it $^\star$Departamento de F\'{\i}sica, 
Universidad de Santiago de Chile,}}\\
{\small {\it Casilla 307, Santiago 2, Chile}}
\\
{\small {\it $^\dagger$Institute for High Energy Physics, Protvino, Russia}}
\\
{\small {\it $^\sharp$E-mail: mplyushc@lauca.usach.cl}}
}
\date{}
 
\maketitle

\begin{abstract}
Deformed Heisenberg algebra  with reflection appeared
in the context of Wigner's generalized quantization schemes  
underlying the concept of parafields
and parastatistics of Green, Volkov, Greenberg and Messiah. 
We review the application of this algebra for the universal description 
of ordinary spin-$j$ and anyon fields in 2+1 dimensions, 
and discuss the intimate relation between parastatistics and supersymmetry.
\end{abstract}

\subsection*{Introduction}
Generalized statistics  
was introduced in physics
in the form of parastatistics as an exotic possibility 
extending the Bose and Fermi statistics 
\cite{pa1,pa2,pa3,pa4,pa5}.  
It was closely related with the discovery
of color in the context of the theory of strong interactions.
Nowadays generalized statistics \cite{an1,an2,an3,an4} finds applications
in the physics of the quantum Hall effect
and (probably) it is relevant to high temperature
superconductivity \cite{an5}.
Supersymmetry, instead, unifies Bose and Fermi statistics 
\cite{su1,su2,su3,su4,su5}
and its development lead to the construction of field
and string theories with exceptional properties 
\cite{su6},
that transformed the same idea of supersymmetry
in one of the cornerstones of modern theoretical physics.
Supersymmetry was observed
in nuclear physics in the form of dynamical symmetry \cite{n1,n2},
whereas its manifestation as a fundamental symmetry
of elementary particle physics still waits for experimental confirmation.

Though supersymmetry and generalized
statistics may be unified in the form of parasupersymmetry \cite{pasu},
nevertheless, by the construction, the two concepts seem to be
independent. Recently, the existence of 
intimate relation between generalized statistics and supersymmetry
was established by observation of hidden supersymmetric structure
in purely parabosonic \cite{hid} and purely pafermionic \cite{kp} systems.
The key tool with which the observation of close relationship
between generalized statistics and supersymmetry was realized
is the so called deformed Heisenberg algebra with reflection, 
or the R-deformed Heisenberg algebra (RDHA) \cite{def1,def2}.
The algebraic construction of RDHA
appeared in the work of Wigner \cite{wig}, 
where he investigated the problem of correlation of
equations of motion with quantum mechanical
commutation relations and proposed the
generalized quantization schemes which
subsequently lead to the 
theoretical discovery of parastatistics \cite{pa1,pa2,pa3,pa4,pa5} 
(in this context, see also refs. \cite{yang,sud1,sud2,msud}).
RDHA represents, probably, one of the first examples
of deformation of bosonic harmonic oscillator which,
as it was shown recently,
possesses some universality being also
related to parafermions \cite{def1,def2}, 
to (2+1)-dimensional anyons \cite{defan1,defan2,defan3},
and to the bosonized form of supersymmetric
quantum mechanics \cite{subos1,defan2,subos2}. 
Besides, the RDHA structure underlies 
the construction of fractional 
supersymmetry  \cite{fsusy}.
In this talk we review the application of RDHA 
for the universal description of ordinary
spin-$j$ and anyon fields in 2+1 dimensions by means of first
order linear differential equations \cite{defan1,defan2,defan3},
and discuss the exotic supersymmetry 
of purely parabosonic systems \cite{hid}.
 
\subsection*{Supersymmetry of parabosons}
The deformed Heisenberg algebra with reflection is
generated by the creation-annihilation
operators $a^+$, $a^-$ and by the reflection operator
$R$ satisfying the relations
\begin{equation}
\label{rdha}
[a^-,a^+]=1+\nu R,\quad \{R,a^\pm\}=0,\quad R^2=1,
\end{equation}
where $\nu$ is a real deformation parameter.
Due to these basic relations, the creation-annihilation
operators satisfy the trilinear
parabosonic commutation relations 
$[\{a^-,a^+\},a^\pm]=\pm 2a^\pm$.
Introducing the Fock vacuum state,
$a^-\vert 0\rangle=0$, and fixing the 
action of operator $R$
on it as $R\vert 0\rangle=\vert 0\rangle$,
we arrive at the relation 
$a^-a^+\vert 0\rangle=(1+\nu)\vert 0\rangle$,
which together with trilinear commutation
relation means that at $\nu=p-1$, $p=1,2,\ldots$,
the operators $a^\pm$ have the sense of single-mode
creation-annihilation operators of paraboson of order 
$p$.
Vice versa, one can show that parabosonic
trilinear commutation relations themselves give rise 
to RDHA \cite{def1}. 
In general case, the number operator is realized in the form  
$N=\frac{1}{2}\{a^-,a^+\}-\frac{1}{2}(1+\nu)$,
and the reflection operator
can be represented as $R=(-1)^N=\cos \pi N$.
 
The reflection operator introduces a natural
$Z_2$ grading structure in the Fock space 
and its presence in the definition of RDHA
can be considered as an indication
on possible relationship between parabosons
and supersymmetry.
To reveal a supersymmetry of parabosons, one notes 
that RDHA can also be given by the relations \cite{hid}
\begin{equation}
\label{f}
a^+a^-=F(N),\quad 
a^-a^+=F(N+1),\quad 
[N,a^\pm,]=\pm a^\pm,
\end{equation}
where
$F(N)=N+\nu\sin^2\frac{\pi N}{2}$
is the characteristic function satisfying for
$\nu>-1$ the relations $F(0)=0$, $F(n)>0$,
$n=1,2,\ldots$. These relations mean, in particular,
that for $\nu>-1$ the corresponding representations
of RDHA are unitary and infinite-dimensional.
On the other hand, in the case $\nu=-(2p+1)$, $p=1,2,\ldots$,
the characteristic function possesses the property
$F(2p+1)=0$ underlying the existence of $(2p+1)$-dimensional
(non-unitary) irreducible representations of RDHA,
which are associated with the
deformed parafermions of order $2p=2,4,\ldots$ \cite{def1,def2,kp}.
Let us restrict ourselves here by the case 
of unitary infinite-dimensional representations
($\nu>-1$). 
When $\nu=2k+1$, $k=0,1,\ldots$,
the structure function
satisfies the relation
$F(2n+1)=F(2n+\nu+1)$, $n=0,1,\ldots$.
Therefore, in the case of parabosonic systems of even order
$p=2(k+1)$, the spectrum of the
quadratic Hamiltonian $H=a^+a^-$ (or of $H=a^-a^+$),
reveals doubling of all higher-lying levels.
This indicates on existence of supersymmetry
in such purely parabosonic systems. 
As it follows from the explicit form
of the characteristic function,
in the case of paraboson of order $p=2(k+1)$, $k=0,1,2\ldots$,
and Hamiltonian $H=a^+a^-$, 
the supersymmetry  is characterized by the presence of 
$k+1$ singlet states with energies $E=0,2,\ldots,2k$.
Therefore, only in the case $k=0$, the spectrum has one 
singlet state of zero energy, whereas all other cases
are characterized by the presence of $k$ higher-lying 
singlet states of nonzero energy
in addition to the zero energy ground state. 
The corresponding supercharges are the infinite series
operators in the corresponding paraboson operators:
\begin{equation}
\label{q}
Q_+=(a^+)^{2k+1}\sin^2 \pi J_0,\quad Q_-=(a^-)^{2k+1}\cos^2 \pi J_0,\quad
J_0=\frac{1}{4}\{a^+,a^-\}.
\end{equation}
They together with the Hamiltonian satisfy the polynomial
superalgebra: 
\begin{eqnarray}
&Q_\pm^2=0,\quad 
[H,Q_\pm]=0,&\nonumber\\
&\{Q_+,Q_-\}=(H-2k)(H-2k+2)\ldots(H+2k-2)(H+2k),&
\label{poly}
\end{eqnarray}
which in the case $k=0$ is reduced to the
conventional $N=1$ linear superalgebra.
The role of the grading operator in such purely
parabosonic supersymmetric systems belongs to
the reflection operator $R=\cos\pi N$.
 
The case of paraboson system of order $p=2$ ($\nu=1$)
given by the Hamiltonian $H=a^-a^+$ is characterized
by the $N=1$ spontaneously broken 
linear supersymmetry:
all the states are paired in supersymmetric doublets
with the lowest  energy level $E=2$. The 
corresponding supercharges have the form (\ref{q})
with $k=0$ and with
operators $a^+$ and $a^-$ changed in their places \cite{subos1}.
The systems of parabosons of order $p=4,6,\ldots$, 
given by the Hamiltonian $H=a^-a^+-2$
possess nonlinear (polynomial) supersymmetry
of the form similar to (\ref{poly}) \cite{hid}.

It was shown in ref. \cite{hid}
that the supersymmetry of purely parabosonic systems
can be understood as the supersymmetry of
Calogero-like systems with exchange interaction 
and that in principle it can be realized 
in one-dimensional systems
of identical fermions. Besides, it was demonstrated that 
nonlinear parabosonic supersymmetry can be obtained
via appropriate modification of the classical analog
of usual supersymmetric quantum mechanics.

\subsection*{RDHA and anyons}
The parabosonic supersymmetry
structures corresponding to $H=a^+a^-$ and $H=a^-a^+$
can be unified and extended to
the $osp(2\vert2)$ superalgebraic structure \cite{subos1}. 
The operators $T_3=\frac{1}{2}\{a^+,a^-\}$,
$T_\pm=\frac{1}{2}(a^\pm)^2$ and 
$I=\frac{1}{2}(\nu+R)$ have a sense of even generators
of $osp(2\vert 2)$ forming $sl(2)\times u(1)$ subalgebra,
whereas the operators $Q^\pm=a^\pm\Pi_\pm$
and $S^\pm=a^\pm\Pi_\mp$ are its odd generators, 
where $\Pi_\pm=\frac{1}{2}(1\pm R)$
are the projectors on even and odd subspaces
of the Fock space.
On the other hand, the operators $J_\mu$,
$\mu=0,1,2$,
$J_0=\frac{1}{2}T_3$, 
$J_1\pm iJ_2=T_\pm$,
and ${\cal L}_\alpha$,
$\alpha=1,2$,
${\cal L}_1=(a^++a^-)/\sqrt{2}$,
${\cal L}_2=i(a^+-a^-)/\sqrt{2}$, 
can be considered as
even and odd generators of $osp(1\vert 2)$
superalgebra:
$[J_\mu,J_\nu]=-i\epsilon_{\mu\nu\lambda}J^\lambda$, 
$[J_\mu,{\cal L}_\alpha]=
\frac{1}{2}(\gamma_\mu)_\alpha{}^\beta{\cal L}_\beta$,
$\{{\cal L}_\alpha,{\cal L}_{\beta}\}=4i(J\gamma)_{\alpha\beta}$;
here (2+1)-dimensional $\gamma$-matrices 
appear in the Majorana representation, 
see ref. \cite{defan1}.
Since the Casimir operator ${\cal C}=J_\mu J^\mu-
\frac{i}{8}{\cal L}^\alpha{\cal L}_\alpha$
takes the fixed value
${\cal C}=\frac{1}{16}(1-\nu^2)$,
this means that any irreducible representation
of RDHA carries the corresponding irreducible
representation of $osp(1\vert 2)$,
which, in turn, is a direct sum
of two irreducible representations
of $so(2,1)$
with the Casimir operator $C=J^\mu J_\mu$
taking the value 
$C=-\alpha_{+}(\alpha_{+}-1)$, 
$\alpha_+=\frac{1}{4}(1+\nu )$,
on even ($R=1$) subspace of the Fock space,
and $C=-\alpha_-(\alpha_--1)$,
$\alpha_-=\alpha_++\frac{1}{2}$,
on odd ($R=-1$) subspace.
The $osp(1\vert 2)$ structure associated with
RDHA can be exploited to describe anyons
by means of covariant linear differential equations.
For the purpose, let us consider 
the field $\Psi^n(x)$
depending on space-time point $x_\mu$ in 2+1 dimensions
and carrying infinite- ($\nu>-1$, $n=0,1,\ldots$) 
or finite- dimensional
($\nu=-(2p+1)$, $p=1,2,\ldots$, $n=0,\ldots,2p$) 
representation
of RDHA. Then the following spinor set of linear
differential equations describes universally ordinary
spin-$j$ fields and anyons \cite{defan1,defan2,defan3}:
\begin{equation}
\label{sd}
D_\alpha \Psi(x)=0,\quad
D_\alpha=R{\cal P}_\alpha+m{\cal L}_\alpha,
\end{equation}
where $R$ is the reflection operator of RDHA,
$m$ is a mass parameter, 
and ${\cal P}_\alpha=
(-i\gamma_\mu\partial^\mu)_\alpha{}^\beta{\cal L}_{\beta}$.
The condition of integrability of two equations (\ref{sd})
is equivalent to the equations $(-\partial^2+m^2)\Psi_+(x)=0$,
$(-i\partial^\mu J_\mu-sm)\Psi_+(x)=0$, $s=\frac{1}{4}(1+\nu)$,
for the even part (in the RDHA sense) of the field, $R\Psi_+(x)=\Psi_+(x)$,
whereas the solution to equations (\ref{sd})
in odd subspace,  $R\Psi_-(x)=-\Psi_-(x)$, 
is trivial, $\Psi_-(x)=0$.
The parameter $\nu$ fixes the value of spin $s$, 
and one concludes that in the case of finite-dimensional
representations ($\nu=-(2p+1)$),
the corresponding field $\Psi_+(x)$ carries
integer or half-integer spin $s=j$, 
whereas the case of
infinite-dimensional representations of RDHA
($\nu>-1$) corresponds to the field of arbitrary spin
$s>0$ (anyon). The case of anyon with $s<0$ can be obtained
by a simple change $m\rightarrow -m$ in (\ref{sd}).
In the case of infinite-dimensional unitary representations
of RDHA, the linear differential
equation $(-i\partial^\mu J_\mu-sm)\Psi_+(x)=0$
is the (2+1)-dimensional analog of 
the $(3+1)D$ infinite-component Majorana equation,
whose fundamental role for the description of anyons
was established in ref. \cite{tor} under investigation
of the $(2+1)D$ model of relativistic
particle with torsion (see also refs. \cite{jn,dmaj}). 
 
Varying the deformation
parameter in the region $\nu>-1$,
one can obtain  the fields of integer spin ($\nu=4n-1$, $s=n$, $n=1,2,\ldots$)
as well as of half-integer spin ($\nu=4n+1$, $s=n+\frac{1}{2}$,
$n=0,1,\ldots$). However, such fields of integer and half-integer spin have 
a nature to be essentially different from the nature
of usual spin-$j$ fields appearing in the case of
finite-dimensional representations of RDHA ($\nu=-(2p+1)$)
since they have hidden nonlocality.
In the rest frame, the solution to the Klein-Gordon and
Majorana equations has only one nontrivial component
in correspondence with the pseudoscalar (helicity) nature of spin 
in 2+1 dimensions \cite{corp,mono}. 
But the Lorentz boost 
enlivens all the infinite number of components
of the field $\Psi_+(x)$ in the case $\nu>-1$ \cite{defan3}.
There is the analog of coordinate representation
for RDHA, in which 
\begin{equation}
\label{yd}
a^\pm=\frac{1}{\sqrt{2}}(q\mp i{\cal D}_{\nu}),\quad
{\cal D}_\nu=-i\left(\frac{d}{dq}-\frac{\nu}{2q}R\right),
\end{equation}
and $R\psi(q)=\psi(-q)$.
In such representation the fields $\Psi_+(x)$ 
have a structure of the functions to be even in 
continuous variable $q$: $\Psi_+(x,q)=\Psi_+(x,-q)$.
This means that the corresponding solutions to the equations
(\ref{sd}) in the case $\nu>-1$ have the hidden half-infinite
nonlocality ($q\geq 0$) which is analogous to 
the string-like nonlocality of anyon fields 
in other approaches \cite{nonl1,nonl2}.
 
To conclude, RDHA finds various theoretical applications
including the described two.
It would be interesting to look for the experimental
manifestation of the exotic supersymmetry of parabosons.
In general, the existence of the polynomial
supersymmetry is characterized by the presence of several
singlet states which could be considered 
as an indication on parabosonic-like excitations (quasiparticles)
in the system. 
The search for possible experimental manifestation of the nonlocal
(in internal variable $q$)
fields of integer and half-integer spin
associated with parabosons seems to be another interesting 
problem.
 
$\ $
 
\paragraph*{Acknowledgments.} 
I am grateful to I. Bandos, M. Rausch de Traubenberg and D. Sorokin
for useful discussions, and to G. Marmo and 
E. C. G. Sudarshan for bringing 
refs. \cite{sud1,sud2,msud} to my attention.
The work was supported in part by FONDECYT (Chile) under grant 1980619
and by DICYT (USACH).

\end{document}